\documentclass[12pt]{article}

\usepackage{times}
\usepackage{physics}
\usepackage{graphicx}
\usepackage{amssymb}
\usepackage{amsthm}
\usepackage{amsmath}
\usepackage{dsfont}
\usepackage{bm}
\usepackage{mathrsfs}
\usepackage{bbold}
\usepackage{color}
\usepackage[superscript]{cite}

\begin{document}

\section*{Backreaction of stimulated Hawking radiation in an optical analogue}

Lorenzo M. Procopio\footnote{Department of Physics, Paderborn University, Warburger Str. 100, 33098 Paderborn, Germany \& Institute for Photonic Quantum Systems (PhoQS), Paderborn University, Warburger Str. 100, 33098 Paderborn, Germany} \footnote{Department of Physics of Complex Systems, Weizmann Institute of Science, 7610001 Rehovot, Israel}, Raul Aguero-Santacruz$^3$, David Bermudez\footnote{Department of Physics, Cinvestav, A.P. 14-740, 07000 Ciudad de Mexico, Mexico}, and Ulf Leonhardt$^2$\\

\noindent
{\bf 
Hawking radiation \cite{Hawking} --- the emission of quantum particles at the event horizon of a black hole \cite{Brout} --- connects gravity with quantum mechanics and thermodynamics \cite{Bekenstein,Helfer,Polchinski}. But Hawking radiation has never been observed in astronomy, only in laboratory analogues \cite{Weinfurtner,Euve,Steinhauer,Drori} and the chances of ever observing it in space are astronomically small \cite{Drori}. The energy of Hawking radiation must come from the gravitational field around the black hole \cite{Brout}, but how field quanta generate Hawking quanta has been unknown. Here we report on experimental and theoretical evidence for the process that generates Hawking radiation in a fibre--optical analogue of the event horizon \cite{Philbin,Agullo}. There, as in gravity \cite{Brout}, it has been believed that Hawking radiation comes from a complicated, cascaded process \cite{Webb}; here we have identified theoretically a simple, direct process and observed experimentally how this process reacts back onto the field. Our findings suggest an equally direct process for other laboratory analogues \cite{Weinfurtner,Euve,Steinhauer,Nguyen,Viermann,Stein,Shi,Svancara} and perhaps also for gravitational fields, shedding light on how black holes might radiate. 
}

Laboratory analogues \cite{Weinfurtner,Euve,Steinhauer,Drori,Philbin,Agullo,Webb,Nguyen,Viermann,Stein,Shi,Svancara,Volovik,Barcelo,Unsch,Faccio,Jacquet,Unruh} are based on a simple yet powerful idea \cite{Unruh}: imagine waves --- sound waves or light waves --- in a medium moving with variable speed or having variable wave velocity. Suppose the medium moves slower than the waves. In this case  they can travel freely. Suppose now the flow velocity begins to exceed the wave velocity: waves are swept along and trapped. The area where the flow speed exactly equals the wave velocity establishes the analogue of the event horizon. Note that this is not just an analogy, but an exact mathematical equivalence, as the propagation equation for waves in moving media is equivalent to the scalar wave equation in general relativity \cite{Barcelo}. In fibre optics \cite{Agrawal}, light pulses modify the speed of light in the fibre due to the Kerr effect \cite{Agrawal}: the refractive index $n$ acquires a small additional contribution $\delta n$ being proportional to the intensity and moving with the group velocity $u$ of the pulse. In a frame co--moving with the pulse, the glass of the fibre appears as a medium moving with velocity $-u$ while the speed of light $c$ is reduced by $n+\delta n$. For the analogy with gravity we thus need to consider the co--moving frame. Let us give the mathematically simplest argument for the generation of Hawking radiation there (as illustrated in Fig.~1).

%%%
\begin{figure}[h]
\begin{center}
\includegraphics[width=14pc]{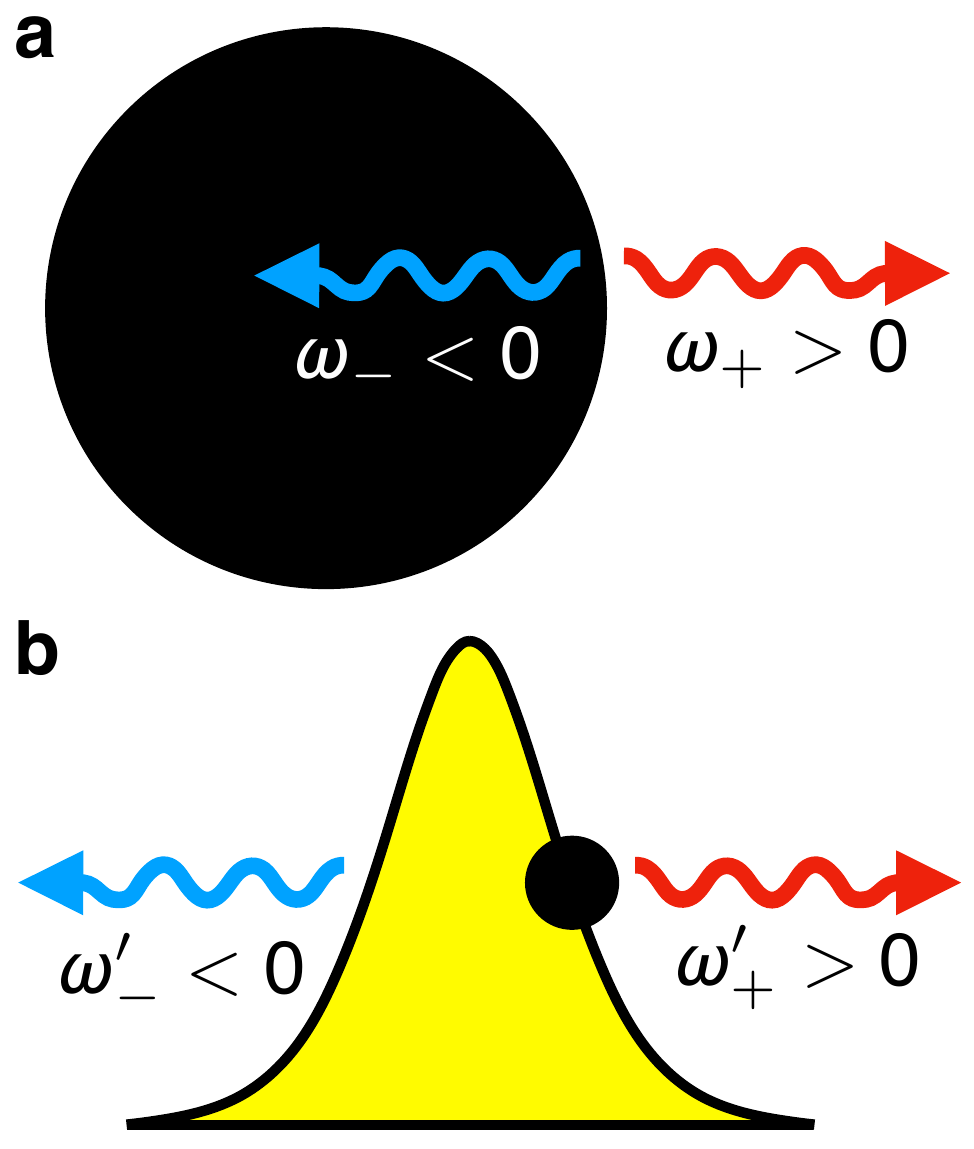}
\caption{
\small{Analogue of the event horizon. {\bf a}: Schematic diagram illustrating the Hawking radiation of an astrophysical black hole. Pairs of quanta are emitted from the event horizon: quanta of positive--frequency waves (indicated in red, with frequency $\omega_+$) escape into space, while their negative--frequency partners (indicated in blue, with frequency $\omega_-$) fall into the singularity of the black hole. The energy for creating the radiation must come from the energy of the gravitational field around the black hole in a process as yet unknown. {\bf b}: Fibre--optical analogue of the event horizon \cite{Philbin}. A light pulse (indicated in yellow) in a fibre changes the local refractive index, acting as a moving medium in a frame co--moving with the pulse. Its front establishes the analogue of a black--hole horizon for probe light (waves) at the point (black dot) where the group velocity of the incident probe (red wave) matches the speed of the pulse. The probe incident with positive co--moving frequency $\omega_+'$ stimulates a wave (red) with the same $\omega_+'$ and another wave (blue) with negative co--moving frequency $\omega_-'$, the analogue of the Hawking partner. Due to the Doppler effect (Fig.~2) the Hawking partner appears in the UV ($233\,\mathrm{nm}$ wavelength, Fig.~3) for probes in the IR ($1100\,\mathrm{nm}$ - $1600\,\mathrm{nm}$). In nonlinear fibre optics \cite{Agrawal} all elementary processes are known in principle. We identify theoretically which ones are responsible for creating Hawking radiation and measure experimentally both the stimulated Hawking radiation and its backreaction (Fig.~4). .  
}
\label{fig:analogy}}
\end{center}
\end{figure}
%%%

Suppose a weak probe pulse interacts with the strong pulse establishing the moving medium, called pump pulse. In fibre optics \cite{Agrawal}, the co--moving frame is defined by the retarded time $\tau =t-z/u$ and the propagation time $\zeta=z/u$ where $z$ denotes the propagation distance in the fibre and $t$ the laboratory time. The frequency $\omega'$ in the co--moving frame is related to the frequency $\omega$ in the laboratory frame by the Doppler effect. For probe light with phase $\varphi$ we obtain from $\omega'=-\partial\varphi/\partial\zeta$, $\omega=-\partial\varphi/\partial t$ and $k=\partial\varphi/\partial z$ with $k=n\omega/c$ the Doppler formula \cite{Philbin}:
%%%
\begin{equation}
\omega' = \left(1-n \frac{u}{c}\right)\omega \,.
\label{doppler}
\end{equation}
%%% 
As the refractive index $n$ depends on $\omega$ the Doppler shift varies with frequency (Fig.~2). Note that the Doppler--shifted frequency is negative when the speed of the moving medium exceeds the phase velocity of light ($u>c/n$). For gravitational black holes \cite{Brout}, Hawking radiation consists of particle pairs, one corresponding to a wave with positive frequency escaping into space and the other to a wave of negative frequency falling into the black hole (Fig.~1). In fibre optics \cite{Philbin}, the Hawking quanta are pairs with frequencies $\omega'$ and $-\omega'$. In our case (Fig.~2) the positive--frequency Hawking waves appear in the laboratory frame  in the infrared (IR, with wavelengths $1100\,\mathrm{nm}$ to $1860\,\mathrm{nm}$) while their negative--frequency Hawking partners are in the ultraviolet (UV, around $233\mathrm{nm}$). 

%%%
\begin{figure}[h]
\begin{center}
\includegraphics[width=32pc]{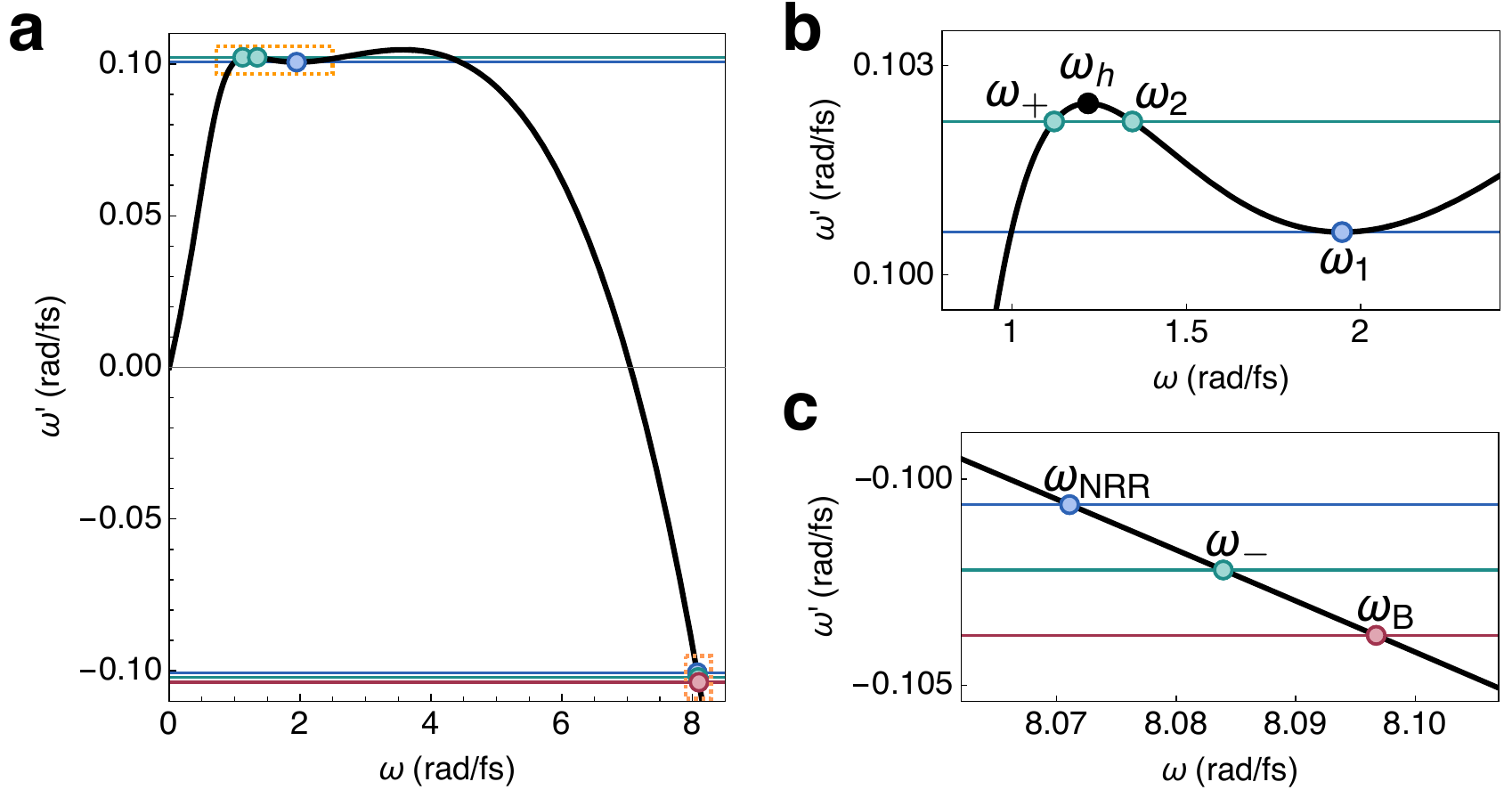}
\caption{
\small{Doppler shift. Co--moving frequency $\omega'$ versus laboratory--frame frequency $\omega$ according to the Doppler formula (\ref{doppler}) with  $n(\omega)$ measured and modeled for the fibre used and $u/c$ fitted to match the measured Hawking frequency (Fig.~3). In the UV the Doppler--shifted co--moving frequency is negative, because there the velocity $u$ of the effective moving medium (Fig.~1b) exceeds the phase velocity $c/n(\omega)$ of the probe. Subfigure {\bf a} shows the entire Doppler curve while {\bf b} and {\bf c} focus on the IR and the UV regions relevant in our experiment. {\bf b}: The light of the pump pulse with frequency $\omega_1$ lies in a local minimum of the Doppler curve, as $\mathrm{d}\omega'/\mathrm{d}\omega = 1-u/v$ with group velocity $v$ and $u=v$ for the pump, while the horizon $\omega_h$ of the probe lies at a local maximum. The probe with incident $\omega_2$ and corresponding $\omega_2'$ gets red--shifted \cite{Philbin} along the Doppler curve over the horizon to $\omega_+$ with $\omega_+'=\omega_2'$. {\bf c}: In the UV the pump pulse generates Negative--frequency Resonant Radiation \cite{Rubino} with co--moving frequency $\omega_\mathrm{NRR}'=-\omega_1'$. The probe stimulates a Hawking partner \cite{Drori} with the negative co--moving frequency $\omega_-'=-\omega_+'=-\omega_2'$. We show that, as backreaction to the Hawking process, the pump acquires a wave with co--moving frequency $\omega_\mathrm{B}'=-\omega_2'-(\omega_2'-\omega_1')$ that we observe (Figs.~3 and 4) at the frequency $\omega_\mathrm{B}$ further in the UV. 
}
\label{fig:analogy}}
\end{center}
\end{figure}
%%%

Let us derive the dynamics of the light in the co--moving frame. We describe the optical field by the dimensionless amplitudes $A_1$ for the pump and $A_2$ for the probe. Pump and probe are distinct light fields of different frequency range and possibly different polarization. In our case, they are separated in frequency $\omega$ by about an octave but share the same polarization (for maximizing their interaction \cite{Agrawal}). Classical fields $A_m$ consist of the analytic signals \cite{Amiranashvili1,Amiranashvili2,Zakharov} $a_m$ as $A_m=\frac{1}{2}(a_m+a_m^*)$ where the $a_m$ oscillate with positive laboratory frequencies \cite{Aguero}. For quantum fields the $a_m$ correspond to the local annihilation operator $\widehat{a}_m$ and the $a_m^*$ to its Hermitian conjugate. Writing $a_m$ in terms of quadratures \cite{Leonhardt} as $a_m=2^{-1/2}(q_m+\mathrm{i}p_m)$ we obtain from Hamilton's equations $\dot{q}_m = \partial H/\partial p_m$ and $\dot{p}_m = -\partial H/\partial q_m$ the propagation equation 
\begin{equation}
\mathrm{i}\dot{a}_m = \frac{\partial H}{\partial a_m^*} 
\label{propagation}
\end{equation}
%%%  
where the dot denotes differentiation with respect to propagation time $\zeta$. The Hamiltonian \cite{Amiranashvili1,Zakharov} $H$ has units of frequency here and it consists of three parts, $H=H_1+H_2+H_\mathrm{int}$, the independent--propagation Hamiltonians $H_m$ of each field and the interaction $H_\mathrm{int}$ (Methods Sec.~A). In nonlinear fibre optics \cite{Agrawal}, the interaction is due to the Kerr effect \cite{Agrawal}: $H_\mathrm{int}=16\kappa A_1^2A_2^2$ with coupling constant $\kappa$. Expanding the $A_m$ in terms of the $a_m$ and $a_m^*$ gives all the elementary interactions, and among them the processes described by the Hermitian Hamiltonians:
\begin{equation}
H_\mathrm{R} = 2\kappa\, a_1^*a_1 \left(a_2^{*2}+a_2^2\right) \,,\quad H_\mathrm{S} = 4\kappa\, a_1^*a_1 a_2^*a_2 \,.
\label{hamiltonians}
\end{equation}
%%%  
In the propagation equation (\ref{propagation}) for the probe amplitude $a_2$ the process $H_\mathrm{S}$ adds a contribution linear in $a_2$ and proportional to the pump intensity $a_1^*a_1$. This corresponds to the shift $\delta n$ in the refractive index known in fibre optics as the cross phase modulation \cite{Agrawal}. This shift in refractive index is essential for establishing horizons \cite{Philbin}. 

But the refractive--index shifting $H_\mathrm{S}$  is not the process generating Hawking radiation: that is $H_\mathrm{R}$. To see this, consider a probe wave incident with co--moving frequency $\omega_2'$ and interacting with a pump wave of frequency $\omega_1'$. In Eq.~(\ref{propagation}) for $a_2$ the Hamiltonian $H_\mathrm{R}$ generates a term oscillating with $-\omega_2'$. A probe wave of positive co--moving frequency $\omega_2'$ thus stimulates a wave with negative $-\omega_2'$, the Hawking partner \cite{Brout} (Fig.~1). In the limit of zero average probe intensity, we need to replace the amplitudes $a_m$ and $a_m^*$ by quantum amplitudes $\widehat{a}_m$ and $\widehat{a}_m^\dagger$. The quantum Hamiltonian should be normally ordered \cite{Leonhardt}, because otherwise it would create radiation even if the pump were in the vacuum state. For an intense pump reaching the classical limit,  $H_\mathrm{R}$ of Eq.~(\ref{hamiltonians}) is a standard pair--creation Hamiltonian \cite{Leonhardt}. It  spontaneously generates an entangled two--mode squeezed state \cite{Leonhardt} of Hawking quanta from the quantum vacuum,  one with positive and the other with negative co--moving frequency. This shows that $H_\mathrm{R}$ is responsible for creating Hawking radiation. In our experiment, we do not yet observe the fully quantum spontaneous Hawking radiation, but classical stimulated Hawking radiation. While this classical radiation shares many of the features of quantum Hawking radiation, in particular its spectrum, we would not be able to measure quantum entanglement.

Note that in the experiment $H_\mathrm{R}$ requires a pump pulse with near--single--cycle features and hence ultrabroad spectrum, as pump and probe are not frequency matched (Methods Sec.~B). For this reason, $H_\mathrm{R}$ is normally neglected in nonlinear fibre optics\cite{Agrawal}. One notable exception is Negative--frequency Resonant Radiation \cite{Rubino} (NRR, Fig.~2c) that was predicted  \cite{Rubino} and observed \cite{Drori,Rubino} as the result of a similar process \cite{Conforti,Aguero}. The Hamiltonian $H_\mathrm{R}$ was  also neglected  in the previous discussions of fibre--optical event horizons \cite{Webb}. But there it is essential: instead of a complicated, cascaded process \cite{Webb}, Hawking radiation is made directly by the simple Hamiltonian $H_\mathrm{R}$. Note also that the generation of Hawking radiation only becomes effective when the probe wave travels with the pump wave, {\it i.e.}\ when it has the same group index as the pump, $c/u=\partial (n\omega)/\partial\omega$ including the change $\delta n$ due to $H_\mathrm{S}$. This is the horizon condition: the optical analogue of the event horizon is established when the group velocity of the probe $A_2$ reaches the speed of the moving medium made by the pump pulse $A_1$. Both aspects of the interaction must act in unison, but the actual pair creation is done by $H_\mathrm{R}$. Note also that in our case the difference between group index $c/u$ and phase index $n$ is an order of magnitude larger than the perturbation $\delta n$ caused by the Kerr effect \cite{Agrawal} of the pump ($\delta n\sim 10^{-3}$). This implies that we are in the regime of strong dispersion. For weak dispersion, Hawking radiation has a thermal spectrum \cite{Hawking,Brout,Bekenstein} with Bekenstein--Hawking temperature \cite{Philbin} $k_\mathrm{B} T =\hbar/(2\pi\tau_0)$ and $\tau_0^{-1}=\partial\ln\delta n/\partial\tau$ evaluated at the horizon. We found (Methods Sec.~B) that for strong dispersion and short pump pulses  $\tau_0$ is the asymptotic exponent of the Fourier transform of $\delta n$ for large frequencies. 
 
We have identified the elementary process generating Hawking radiation in our optical analogue, the Hamiltonian $H_\mathrm{R}$ that describes the transfer of energy between pump and probe, which corresponds to the energy transfer between the gravitational field and Hawking radiation. But as {\it actio = reactio} this Hamiltonian also describes the backreaction of the Hawking radiation onto the pump.  As $H_\mathrm{R}$ is proportional to $a_1^*a_1$ it does not change the total photon number of the $a_1$ field that includes the pump, but redistributes the original pump photons to a different frequency. Given probe and pump with co--moving frequencies $\omega_1'$ and $\omega_2'$ we obtain from Eq.~(\ref{propagation}) for $a_1$ a term oscillating with $-\omega_2'-\delta'$ where $\delta'=\omega_2'-\omega_1'$. As $\omega_2'>\omega_1'$ (Fig.~2) this co--moving frequency is more negative than $-\omega_2'$, which implies (Fig.~2) that the corresponding laboratory frequency is further shifted to the UV (by $0.5\, \mathrm{nm}$ to $1\,\mathrm{nm}$ in wavelength). As the backreaction of Hawking radiation is a secondary process it goes with the square of the probe power. 

Note that our backreaction resembles the backreaction in the Unruh effect \cite{UnruhEffect}. There, an accelerated observer receives the equivalent of Hawking radiation from the quantum vacuum in empty Minkowski space. For an observer at rest, the detection of an Unruh quantum appears as the emission of a Minkowski quantum \cite{UnruhWald}.  Multiple such events slow the acceleration down. In our case, the backreaction plays the role of the emission  \cite{UnruhWald} in Minkowski space while the eventual evaporation of the pump pulse (or the black hole \cite{Hawking}) corresponds to the eventual deceleration in the Unruh effect \cite{UnruhEffect}.

In order to observe the backreaction experimentally, we used the same setup as in our previous observation of stimulated Hawking radiation \cite{Drori}, but improved the stability and the data taking such that we could increase the signal--to--noise ratio to the required level. Both pump and probe pulses are split off from the same train of ultrashort light pulses (of $8\mathrm{fs}$ duration, $800\,\mathrm{nm}$ carrier wavelength, $80\mathrm{MHz}$ repetition rate, from a Thorlabs Octavius laser). The probe undergoes frequency shifting \cite{Rosenberg} in a $1\mathrm{m}$ photonic--crystal fibre (NKT Photonics NL-1.9-765) by the Raman effect \cite{Agrawal}, depending on its intensity that we control. For the pump we use chirped mirrors and a wedge to compensate and control its chirp. Pump and probe are recombined using broad--band optics (parabolic mirrors) and interact in a $7\mathrm{mm}$ piece of photonic--crystal fibre (NKT Photonics NL-1.5-590). From the outgoing light we remove the pump contribution by dichroics and either detect its IR or UV spectrum. In the IR we can observe the frequency shifting due to the analogue of the event horizon \cite{Philbin}, in the UV we count the photons of Hawking partners with negative co--moving frequency \cite{Drori} and the photons of the presumed backreaction. We use a combination of filters to reduce the background noise. For the UV detection, we send the beam via a rotable prism and through a pinhole to a photomultiplier tube (Hamamatsu H8259) for counting the photons of the wavelength $\lambda$ set by the angle of the prism and the pinhole. The apparatus is calibrated with a Mercury lamp. While taking data we continuously monitor the interaction between probe and pump by measuring with a photodiode the frequency shifting of the probe in the IR \cite{Philbin}. A large part of the UV photons comes from Negative--frequency Resonant Radiation  \cite{Rubino} of the pump itself (Fig.~2c). We remove this feature by chopping the train of probe pulses and subtracting from the spectrum with pump--probe interaction the spectrum without interaction (Fig.~3). The pump is easily perturbed by probes with intensities as low as $10^{-2}$ of the pump \cite{Demircan} as the pump is primarily a soliton \cite{Solitonen} that reacts rigidly to perturbations. In particular, the carrier frequency gets shifted (by $10^{-3}$) which results in a shift of the NRR we fit and correct for (Methods Sec.~D). Without the correction, the difference signal may become negative; some remnants are still visible (Fig.~4). 

%%%
\begin{figure}[h]
\begin{center}
\includegraphics[width=35pc]{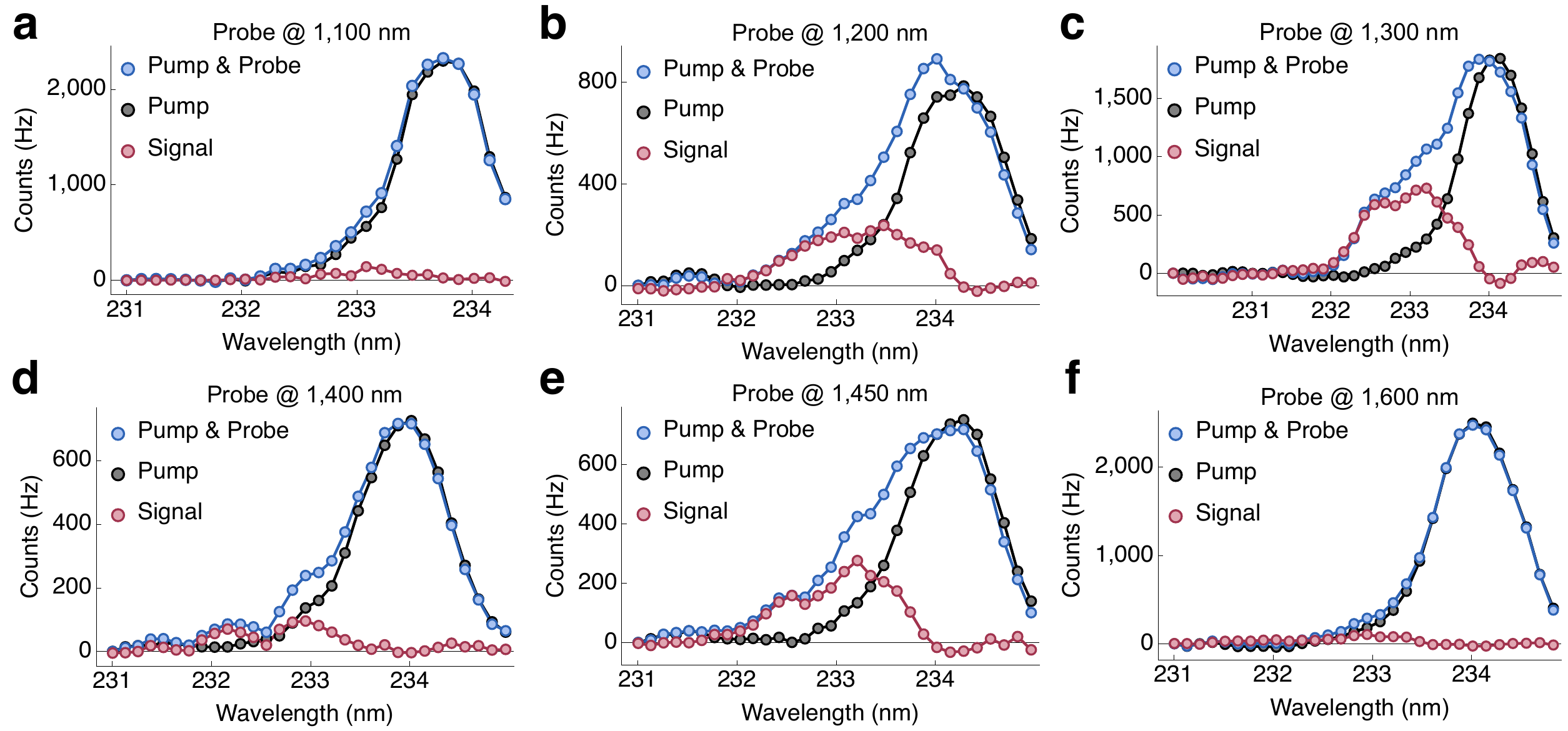}
\caption{
\small{Experimental data. We measure the UV spectra of the pump--probe interaction and the pump beam alone  for several probe wavelengths: 1100 nm, 1200 nm, 1300 nm, 1400 nm, 1450 nm, and 1600 nm, displayed in \textbf{a}-\textbf{f}, respectively. The subfigures show the counts of light quanta per second for bins of wavelength $\lambda=2\pi c/\omega$ (dots). Blue: both pump and probe are present and interact with each other. Black: the probe is chopped out such that the pump interacts only with itself, producing Negative--frequency Resonant Radiation \cite{Rubino} at $\lambda_\mathrm{NRR}$ (Fig.~2c). Red: corrected difference between the counts shown in detail and compared with theory in Fig.~4. We correct the difference by shifting the spectra for the pump by a small frequency we determine by fitting to the pump--probe spectra around and to the right of the main peaks (Methods Sec.~D). Each experiment (except {\bf f}) was independently repeated on another day with consistent results. Statistical errors are smaller than the size of the data points.
}}
\end{center}
\end{figure}
%%%

%%%
\begin{figure}[h]
\begin{center}
\includegraphics[width=35pc]{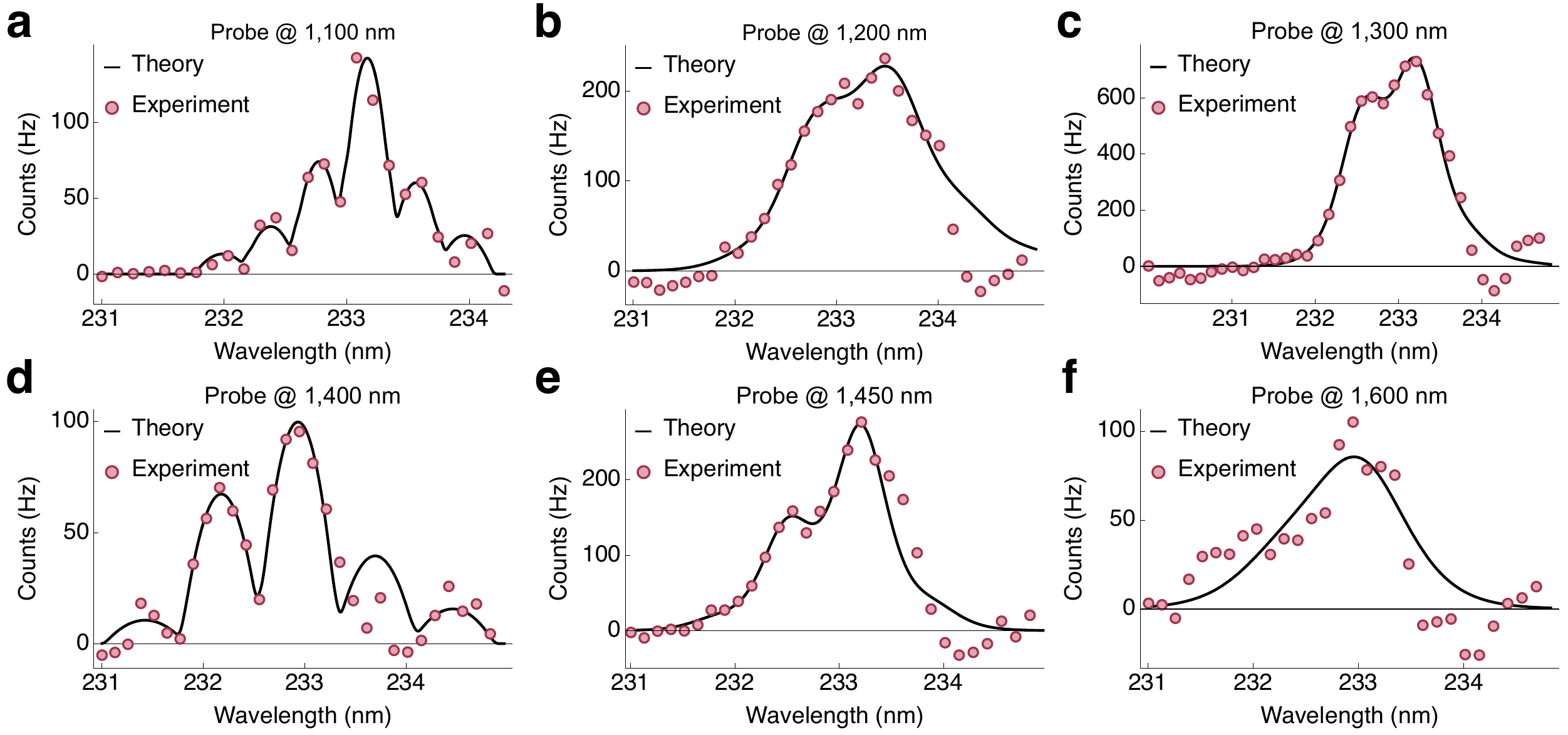}
\caption{
\small{Theory versus experiment. We fit the signals (Fig.~3) with the theoretical curve for each probe wavelength:  1100 nm, 1200 nm, 1300 nm, 1400 nm, 1450 nm, and 1600 nm, shown in \textbf{a}-\textbf{f}, respectively. The circles correspond to the corrected differences in the UV spectra between the pump--probe and pump data (Fig.~3) while the black curves are fits with theory (Methods Sec. C-D). The fits agree well with the data, except in the spectral region where the signal interferes with the Negative--frequency Resonant Radiation \cite{Rubino}. In each subfigure we see a central peak of Hawking radiation and several sidebands (in particular in \textbf{a} and \textbf{d}) due to modulation at the beat frequency of pump and probe. The parameters of the theory curves are fitted to the left and at the Hawking peaks, but in \textbf{a} and \textbf{d} also reproduce the right sidebands, which indicates the consistency of our correction procedure. Without the backreaction, the sidebands were symmetric; their visible asymmetry reveals the backreaction. 
}}
\end{center}
\end{figure}
%%%

In the absence of a systematic study of the signal as a function of probe power that would allow one to unambiguously separate genuine Hawking backreaction from direct probe-induced effects, it remains difficult to exclude the possibility that effects beyond a simple frequency shift contribute to the observed backreaction signal. These may include probe-induced modifications of phase-matching conditions, conversion efficiency, pump dynamics, or additional nonlinear mixing. Such measurements are experimentally very demanding, requiring long acquisition times (up to 12 hours) and stability against slow parameter drifts during the whole measurement run, which are not currently achievable in our setup. Nevertheless, the removal of excess signal near the NRR after applying the spectral shift (Fig.~4), together with the emergence of a sideband structure consistent with our model, suggests that the frequency shift constitutes a significant contribution to the probe--induced backreaction effects.

Our probe wavelengths ($1100\,\mathrm{nm}$, $1200\,\mathrm{nm}$, $1300\,\mathrm{nm}$, $1400\,\mathrm{nm}$, $1450\,\mathrm{nm}$, $1600\, \mathrm{nm}$) span the spectrum from the edge ($1100\mathrm{nm}$) of the capture range of the horizon through the spectral region where the probe is redshifted to the beginning ($1600\, \mathrm{nm}$) of the blueshifting. In all cases we clearly see (Methods Sec.~D) the peaks of the stimulated Hawking radiation and the backreaction at the correct positions (Fig.~2). But we also see sidebands (Fig.~4, in particular 4a and 4d) with co--moving frequencies integer multiples $m$ of $\delta'$ away. They are caused by modulation \cite{Agrawal} with the beat frequency of the pump and the probe, similar to the vibrato on a musical string instrument. Normally, frequency modulation of strength $\chi$ generates a spectrum of Bessel functions $J_m(2\chi)\sim\chi^m/m!$ for $|\chi|\ll 1$ and $m\ge 0$. They go with $\chi^m$  because the generation of the $m$--th sideband requires $m$ excitations. As those are indistinguishable, we need to divide $\chi^m$ by the number $m!$ of all permutations. In our case of extremely short pulses, we have a pure power law $X^{|m|}$ with $X$ being a function of $\chi$ (Methods Sec.~C) as the modulation couples many, distinguishable modes. Note that the sidebands of the Hawking radiation and the backreaction lie exactly on top of each other, apart from a shift by $\delta'$, which causes an asymmetry in the spectrum. This asymmetry is the characteristic feature of the backreaction.

We clearly see the asymmetry (Fig.~4). The Hawking and backreaction intensities for all data sets we retrieve by fitting (Methods Sec.~D). The agreement between the observed asymmetry and the analytic predictions is significant and supports the proposed interpretation, although alternative nonlinear processes producing similar effects cannot be explicitly ruled out. Stimulated Hawking radiation should be linear in the probe intensity \cite{Drori} and the backreaction quadratic. Hence we normalize $I_\mathrm{H}$ with respect to the measured probe intensities (expressed in photon counts) and $I_\mathrm{B}$ with their squares. Given the normalized intensities, we can test whether they establish a thermal spectrum. If so, we expect that that the Hawking temperature is determined by the shortest duration $\tau_0$ of the pump during propagation, but $\tau_0$ varies from day to day due to imperceptible changes. To take this into account, we use the NRR as a thermometer. The NRR is essentially Hawking radiation stimulated by the pump itself and so it should have the same temperature. If this hypothesis and our normalizations are correct, the logarithms of $I_\mathrm{H}/I_\mathrm{NRR}^{\omega_+/\omega_1}$ and $I_\mathrm{B}/I_\mathrm{NRR}^{\omega_+/\omega_1}$ plotted over $\omega_+/\omega_1$ should lie on straight lines with equal slopes. Given the simplicity of our model, it is remarkable how well they do this (Fig.~5). The relative root--mean--squares of the straight--line fits of the Hawking radiation and the backreaction are $0.015$ and $0.010$, respectively.  For the ratios of the  slopes we get $1.02$: the Hawking and backreaction temperatures are equal within the accuracy of the linear fits.

%%%
\begin{figure}[h]
\begin{center}
\includegraphics[width=21pc]{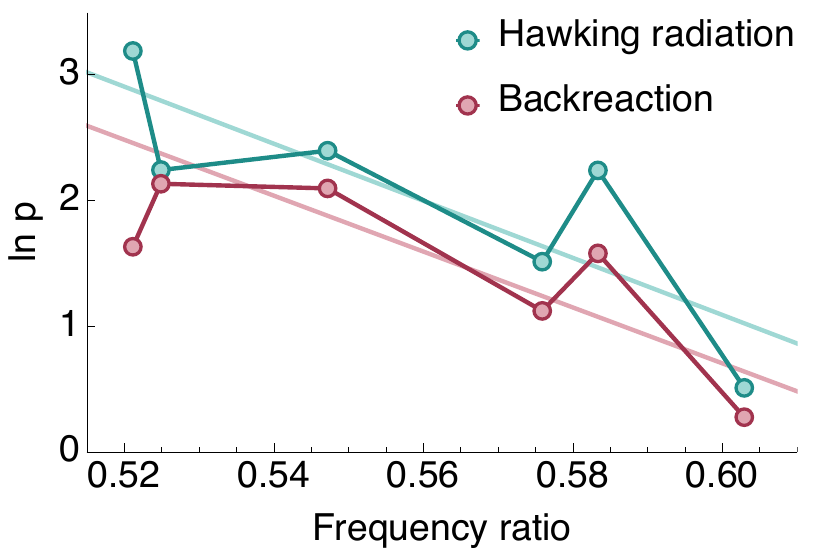}
\caption{
\small{Thermal spectra of Hawking radiation and backreaction. Logarithm of $p_\mathrm{H}=I_\mathrm{H}/I_\mathrm{NRR}^{\omega_+/\omega_1}$ (green dots) and $p_\mathrm{B}=I_\mathrm{B}/I_\mathrm{NRR}^{\omega_+/\omega_1}$ (red dots) for the normalized Hawking radiation and backreaction counts and the NRR counts plotted over $\omega_+/\omega_1=\lambda_1/\lambda_+$ for all six probes in our experiment. For a spectrum with Bekenstein--Hawking temperature \cite{Hawking,Brout,Bekenstein} they would approximate straight lines with equal slopes. They do (lines) and the slopes of those lines agree with $2\%$ accuracy. 
}}
\end{center}
\end{figure}
%%%

We have stimulated Hawking radiation by classical light, but one might use nonclassical light as well, in particular heralded single photons as probes \cite{David}, for testing the entanglement of the Hawking partners. Our findings can be generalized to other analogues of gravity, too. In particular, in Bose--Einstein condensates \cite{Pitaevskii} (BECs) the interaction Hamiltonian is also bi--quadratic (Methods Sec.~E) like in our case of the Kerr interaction in nonlinear fibre optics \cite{Agrawal}. In BECs the backreaction from acoustic horizons was studied theoretically  \cite{Balbinot1,Balbinot2} using a similar phenomenology as in the description of black--hole evaporation \cite{Brout}; the backreaction from suddenly switching on the interaction \cite{Baak} and from the analogue of cosmic inflation \cite{Butera} was studied theoretically, too. The backreaction of water waves on a draining bathtub vortex mimicking a rotating black hole was measured \cite{Patrick} and so was the one on fluids of light \cite{Marino}, but so far all these studies were either theoretical or did not consider the backreaction due to Hawking radiation. 

Our experiment and the underlying theory show that Hawking radiation is the result of a direct process, if the interaction between the radiation and the equivalent of the gravitational field is bi--quadratic. In general relativity, the gravitational field is described by the metric tensor, which can be written in a quadratic form in terms of tetrads \cite{Tetrads} such that the interaction between field and radiation becomes bi--quadratic, as in our case (Methods Sec.~E). Maybe astrophysical black holes radiate by a process as simple and direct as ours. The resulting backreaction would describe in microscopic detail how black holes evaporate, which was the subject of Hawking's 1974 paper \cite{Hawking}. Our experiment also shows that Hawking radiation has a thermal spectrum even in the regime of strong dispersion where the notion of the event horizon loses sense and where the temperature is no longer given by the surface gravity. All this \cite{Maia,Krauss,Susskind} could shed light on the information paradox \cite{Susskind,Calmet}, a problem Hawking struggled with until his very last, 2018 paper \cite{Haco}.

\newpage

\appendix

\small

\renewcommand{\theequation}{\thesection.\arabic{equation}}

\section*{Methods}

\setcounter{equation}{0}

\section{Unidirectional Hamiltonian}

We briefly sketch the Hamiltonian method$^{25,26}$ for nonlinear fibre optics$^{24}$. Our starting point is the one--dimensional wave equation for light with fixed transversal mode structure: $(c^2\partial_z^2-n_0^2\partial_t^2)E=\partial_t^2P_\mathrm{NL}$ where $E$ denotes the electric field strength, $n_0$ the effective refractive index in the fibre and $P_\mathrm{NL}$ the nonlinear polarization$^{24}$. We write the left--hand side as $(c\partial_z+n_0\partial_t) (c\partial_z-n_0\partial_t)E$ and approximate $(c\partial_z-n_0\partial_t)E\approx 2\mathrm{i}n_0\omega E$ and $\partial_t^2P_\mathrm{NL}\approx-\omega^2P_\mathrm{NL}$. Expressing the result in the co--moving coordinates $\tau=t-z/u$ and $\zeta=z/u$ we obtain the unidirectional wave equation$^{51}$:
%%%%%%
\begin{equation}
\left(\mathrm{i}\partial_\zeta-\omega'\right)E = - \frac{\omega u}{2n_0 c}\,P_\mathrm{NL}
\label{eq:wave}
\end{equation}
%%%%%%
with $\omega=\mathrm{i}\partial_\tau$ and $\omega'$ given by Eq.~(1). The nonlinear polarization on the right--hand side may act as a source for new frequencies or it may modulate existing frequencies --- we are going to encounter both cases. The electric field is real and so its Fourier transform $\widetilde{E}$ comprises both positive and negative frequencies. We express the field as $E=a+a^*$ in terms of the analytic signal $a = \frac{1}{\pi} \int_0^\infty \widetilde{E} \,\mathrm{e}^{-\mathrm{i}\omega\tau}\,\mathrm{d}\omega$ that describes the positive laboratory frequencies (while its conjugate accounts for the negative frequencies). Approximating the differential operator $\omega$ on the right--hand side by the carrier frequency we obtain for the analytic signal:
%%%%%%
\begin{equation}
\left(\mathrm{i}\partial_\zeta-\omega'\right)a = \left(\frac{\partial  H_\mathrm{int}}{\partial a^*} \right)_+  \quad\mbox{with}\quad \frac{\partial  H_\mathrm{int}}{\partial a^*} = - \frac{\omega u}{2n_0 c}\,P_\mathrm{NL}
\label{eq:analytic}
\end{equation}
%%%%%%
where the plus sign denotes a projection to positive frequencies and $H_\mathrm{int}$ is the Hamiltonian of the nonlinear interaction. For the Kerr effect$^{24}$ this Hamiltonian is proportional to $E^4$. Expanding $E^4=(a+a^*)^4 = a^4+4a^3a^*+6a^{*2}a^2+4a^{*3}a+a^{*4}$ gives all processes derived from the Kerr nonlinearity. These include the familiar self--phase and cross--phase modulation$^{24}$, third--harmonic generation$^{24}$, the generation of dispersive waves$^{52}$ and supercontinua$^{53}$, but also Negative--frequency Resonant Radiation (NRR)$^{30,31}$ and stimulated Hawking radiation$^9$. Each individual Hermitian sub--Hamiltonian $H_i$ defines, for the corresponding sub--process, a conservation law of the energy in the co--moving frame:
%%%%%%
\begin{equation}
E_i = \int \left(a^*\omega' a + H_i\right)\mathrm{d}\tau = \mathrm{const.}
\label{eq:energy}
\end{equation}
%%%%%%
Indeed, we obtain $\partial_\zeta E_i=0$ from $(\mathrm{i}\partial_\zeta-\omega')a = \partial H_i/\partial a^*$ as $\omega'$ is a Hermitian operator with respect to $\tau$. In our paper, we decompose the light field into $a = a_1 +a_2$ where $a_1$ contains the pump pulse (and new frequencies generated) and $a_2$ the probe (and the results of its interaction with the pump). The Hamiltonian $H_i$ describes then how energy from $a_1$ is converted to energy in $a_2$ and vice versa, {\it actio et reactio}.

\setcounter{equation}{0}

\section{Hawking radiation and backreaction}

Let us derive the creation of stimulated Hawking radiation  and its backreaction from the optical propagation equation (\ref{eq:analytic}) and show in what respect fibre-optical Hawking radiation is thermal. Consider first the sub--Hamiltonian $H_\mathrm{S}\propto a^{*2}a^2\sim a_1^*a_1\,a_2^*a_2$. This Hamiltonian generates for $a_2$ a nonlinear polarization of the form $P_\mathrm{NL}=2n_0\,\delta n\,a_2$ where $\delta n\propto a_1^*a_1$. Comparing this term in Eq.~(\ref{eq:analytic}) with Eq.~(1) for $\omega'$ we see that $\delta n$ is the nonlinear contribution to the refractive index. This process describes the red-- or blue--shifting of the probe at the horizon$^{10}$. For redshifting ($\omega_2>\omega_h$ of Fig.~2) the probe shifts from $\omega_2$ to a lower $\omega_+$ (for $\omega_2<\omega_h$ it blueshifts to a higher frequency) while keeping $\omega_+'=\omega_2'$. The resulting wave is the positive--frequency Hawking radiation (Fig.~1). 

Consider now the process with sub--Hamiltonian $H_\mathrm{R}\propto a^* a (a^{*2}+a^2)\sim a_1^* a_1 (a_2^{*2}+a_2^2)$ generating $P_\mathrm{NL}=2n_0\,\delta n\,a^*_2$. The conjugate of the initial $a_2$ oscillates with negative co--moving frequency $-\omega_2'$ and thus generates the Hawking partner with negative co--moving frequency (Fig.~1). In terms of the laboratory frequency, it corresponds to $\omega_-$ with $\omega'(\omega_-)=-\omega'(\omega_2)$. We will argue that for $\omega_2>\omega_h$ we should take the redshifted probe as initial wave $a_+$. We thus have:
%%%%%%
\begin{equation}
\left(\mathrm{i}\partial_\zeta-\omega'\right)a_- = -\omega_-\frac{u}{c}\,\delta n\, a_+^* \,.
\label{eq:hawkingwave}
\end{equation}
%%%%%%
This is a wave equation in space and time as the left--hand side depends on $\omega=\mathrm{i}\partial_\tau$ via Eq.~(1) while the right--hand side contains the moving index perturbation $\delta n(\tau)$. We Fourier--transform Eq.~(\ref{eq:hawkingwave}) with respect to $\tau$ for the laboratory frequency $\omega_-$. The product $\delta n\, a^*_+$ turns into the convolution $\frac{1}{2\pi}\int_{-\infty}^{+\infty} \widetilde{\delta n}(\omega_--\omega)\,\widetilde{a^*_+}(\omega)\,\mathrm{d}\omega\sim \widetilde{\delta n}(\omega)\,a_+^*(0)$ with $\omega=\omega_++\omega_-$ as the spectrum of $a_+$ is peaked around $\omega_+$ and much narrower than the spectrum of the pump. Note that $a_+^*(0)$ is the complex conjugate of the field $a_+(\tau)$ at the position of the pump ($\tau=0$) and not its Fourier transform. 

Integrating the Fourier--transformed propagation equation (\ref{eq:hawkingwave}) we obtain:
%%%%%%
\begin{equation}
\widetilde{a}_- = \mathrm{i} a_+^*(0)\, \widetilde{\delta n}(\omega)\, \frac{u}{c} \,\omega_-\,\zeta \quad\mbox{with}\quad \omega = \omega_++\omega_- \,.
\label{eq:hawking}
\end{equation}
%%%%%%
Consider, for example, a soliton--like$^{36}$ pump of duration $\tau_0$ where $\delta n=\delta n_0\,\sech^2(\tau/\tau_0)$. In our experiment, the pump undergoes soliton fission$^{53}$ but a sech pulse is still a good model for the most intense fraction. As $\widetilde{\delta n} = \delta n_0\,\pi\omega\tau_0^2\mathrm{csch}(\pi\tau_0\,\omega/2) \sim \delta n_0 \,2\pi \omega \tau_0^2\,\exp(-\pi\tau_0\,\omega/2)$ for large $\omega$ the intensity of the Hawking radiation goes with $\exp(-\pi\tau_0\,\omega_+)$ as a function of the probe frequency $\omega_+$. For $\omega_2>\omega_h$ this frequency is redshifted. The Hawking rate for  the initial $\omega_2$ would go with $\exp(-\pi\tau_0\,\omega_2)$ and  be exponentially lower (as $\omega_2>\omega_+$). Hence it is the redshifted pulse that stimulates the lion's share of the Hawking radiation. For astrophysical black holes, the Hawking radiation is determined by the frequencies of the outgoing light, being redshifted from Planck--scale initial fluctuations, whereas for white--holes it depends on the initial frequencies. This is also the case in our fibre--optical analogue where, for $\omega_2<\omega_h$, the blue--shifted probe contributes exponentially less and therefore we should regard $\omega_+$ as $\omega_2$. Moreover, in either case we can relate the exponent of the optical Hawking rate directly to the astrophysical Hawking temperate $T$. There $T$ is given by the surface gravity$^2$. Translated into optics$^{10}$ $k_\mathrm{B}T=\frac{\hbar}{2\pi}|\partial_\tau\ln\delta n|$ evaluated at the horizon. For a sech pump we get $|\partial_\tau\ln\delta n|=(2/\tau_0)|\tanh(\tau/\tau_0)|$ which gives $k_\mathrm{B}T=\hbar/(\pi\tau_0)$ for large $\delta n$ where the horizon lies in the wings of the sech profile where $\tanh \sim\pm1$. This temperature gives a rate proportional to $\exp(-\hbar\omega_+/k_\mathrm{B}T)$ for large $\omega_+$, which is exactly our rate. Note that for general pulse profiles the optical Hawking temperature is given by the asymptotic exponent $-\pi\tau_0\,\omega$ of the Fourier--transformed $\delta n$ where $\pi\tau_0$ is the lowest pole on the upper complex half plane of $\delta n$ (unless there is no pole as for a Gaussian pulse). This pole and hence the Hawking temperature depends on the entire $\delta n$ profile and not solely on the logarithmic derivative of $\delta n$ at the horizon, which distinguishes our case of weak $\delta n$ and strong dispersion from the ideal astrophysical case. 

We have calculated the negative--frequency Hawking radiation $a_-$ generated by the pump $a_1$ and stimulated by $a_2$. Turn now to its backreaction on $a_1$ with Hamiltonian $H_\mathrm{R}\propto a^* a (a^{*2}+a^2)\sim a_1^* a_1 (a_2^{*2}+a_2^2)$. Equation (\ref{eq:analytic}) with $H_\mathrm{int}=H_\mathrm{R} \propto a_1^*a_1$ implies that the total photon number is conserved, $N_1=\int a_1^* a_1 \,\mathrm{d}\tau = \mathrm{const}$. The pump photons are not lost but some are redistributed within the $a_1$ field. To find out at which frequency they appear differentiate $H_\mathrm{R}\propto a^* a (a^{*2}+a^2)\sim a_1^* a_1 (a_2^{*2}+a_2^2)$ with respect to $a_1^*$. The $a_1 a_2^2$ term oscillates with co--moving frequency $\omega_1'+2\omega_2'$ that lies outside the dispersion curve (Fig.~2a) and so cannot be excited. The $a_1 a_2^{*2}$ term oscillates with $\omega_1'-2\omega_2'=-\omega_2'-\delta'$ where $\delta'=\omega_2'-\omega_1'$ and corresponds to a UV wave of slightly higher frequency $\omega_\mathrm{B}$ than the Hawking radiation (Fig.~2c). For calculating the amplitude $\widetilde{a}_\mathrm{B}$ we may use the same procedure as for the Hawking radiation and obtain:
%%%%%%
\begin{equation}
\widetilde{a}_\mathrm{B}\propto \mathrm{i} a_+^{*2}(0)\, \widetilde{|a_1|}_\omega \,\omega_\mathrm{B}\,\zeta \quad\mbox{with}\quad \omega = \omega_++\omega_\mathrm{B} - (\omega_1-\omega_+) \,.
\label{eq:backreaction}
\end{equation}
%%%%%%
The backreaction is quadratic in the probe intensity, which would make it significantly weaker than the Hawking radiation, but it goes with the rate $\exp(-\pi\tau_0\,\omega)$ for a pump pulse with lowest pole $\pi\tau_0$. This rate is exponentially larger than the Hawking rate, by $\exp[\pi\tau_0(\omega_1-\omega_+)]$ as $\omega_\mathrm{B}\sim\omega_-$ and $\omega_+\approx \omega_1/2$. Consequently, one can observe the backreaction, but there is another complication: overlapping sidebands. 

\setcounter{equation}{0}
\section{Sidebands}

The moving refractive--index perturbation $\delta n$ is proportional to the total intensity $|a_1+a_2|^2\sim|a_1|^2+a_1^*a_2+a_2^*a_1$ for a weak probe. The nonlinear interference term $a_1^*a_2+a_2^*a_1$ oscillates with co--moving frequency $\delta'=\omega_2'-\omega_1'$ and the oscillating $\delta n$ generates sidebands at regular intervals of $\delta'$. Here we work out their amplitudes. 

The sidebands lie at the co--moving frequencies $\omega'_m=\omega_0'+m\delta'$ with integer $m$ around the source at $\omega_0'$. We follow the same procedure as in Sec.~B to derive for their Fourier--transformed amplitudes the equation: 
%%%%%%
\begin{equation}
\left(\mathrm{i}\partial_\zeta-\omega'_m\right)\widetilde{a}_m + \left(2\Omega\cos\delta'\zeta\right) \sum_{n=-\infty}^{+\infty} \widetilde{a}_n = \mathrm{i}\widetilde{g} 
\label{eq:side}
\end{equation}
%%%%%%
where  $\widetilde{g}$ describes the source $\widetilde{g}_-$ of the Hawking radiation or the source $\widetilde{g}_\mathrm{B}$ of the backreaction, respectively. The parameter $\Omega$ is proportional to the Fourier transform of $|a_1^*a_2|$ at $\omega_m-\omega_n$. As $|a_1^*a_2|$ is short in time it has a wide, nearly uniform spectrum that gives the same $\Omega$ for each coupling. For solving Eq.~(\ref{eq:side}) we define ${\cal A}=\widetilde{g} + (2\mathrm{i}\,\Omega\cos\delta'\zeta)\sum_m a_m$ and express ${\cal A}$ in a Fourier series with coefficients ${\cal A}_n$ (assuming that ${\cal A}$ is periodic in $2\pi/\delta'$). One verifies that 
%%%%%%
\begin{equation}
\widetilde{a}_m = \mathrm{e}^{-\mathrm{i}\omega_m'\zeta}\sum_{n=-\infty}^{+\infty} {\cal A}_n\, \mathrm{e}^{\mathrm{i}(\omega_m'-\omega_n')\zeta/2}\,\zeta\, \mathrm{sinc}\left(\frac{\omega_m'-\omega_n'}{2}\,\zeta\right) \sim \mathrm{e}^{-\mathrm{i}\omega_m'\zeta}\,\zeta\,{\cal A}_m 
\label{eq:sidesolution}
\end{equation}
%%%%%%
is an exact solution of $(\partial_\zeta+\mathrm{i}\omega_m')\widetilde{a}_m={\cal A}$ [Eq.~(\ref{eq:side})] with the initial condition $\widetilde{a}_m=0$ for $\zeta=0$. Summing up the $\widetilde{a}_m$ of Eq.~(\ref{eq:sidesolution}) in the definition of ${\cal A}$ gives ${\cal A} =[1+2\pi\mathrm{i}\,(\Omega/\delta')\cos\delta'\zeta]^{-1}\widetilde{g}$ (which is indeed periodic). It remains to determine the Fourier coefficients: 
%%%%%%
\begin{equation}
{\cal A}_m = \widetilde{g}\,\frac{1-X^2}{1+X^2}\,(-\mathrm{i}X)^{|m|} \quad \mbox{with}\quad X = \frac{\sqrt{1+(2\pi\chi)^2}-1}{2\pi\chi} \,,\quad \chi=\frac{\Omega}{\delta'}\,.\label{eq:coeff}
\end{equation}
%%%%%%
The sidebands would obey an exact power law symmetric around the Hawking peak --- were it not for the backreaction. 

The backreaction is generated exactly one sideband to the left of the main Hawking peak (at $\omega_\mathrm{B}'=\omega_-'-\delta'$). Therefore, the bands of  the backreaction align perfectly with the Hawking bands, but are shifted by $-\delta'$. Both bands are coherent; we need to add their amplitudes, and obtain:
%%%%%%
\begin{equation}
\widetilde{a}_m \sim  \widetilde{g}_m\zeta\, \mathrm{e}^{-\mathrm{i}\omega_m'\zeta}\,\frac{1-X^2}{1+X^2}\,(-\mathrm{i}X)^{|m|} \quad\mbox{with}\quad \widetilde{g}_m = \begin{cases}
\widetilde{g}_- + \mathrm{i}\widetilde{g}_\mathrm{B}/X\,\,\,:\,m< 0 \\
\widetilde{g}_- - \mathrm{i}\widetilde{g}_\mathrm{B}X \quad\,:m\ge0 \,.
\end{cases}
\label{eq:sum}
\end{equation}
%%%%%%
We see that the intensities of the total radiation bands around the main Hawking peak are given by simple power laws with asymmetric prefactors. Left of the Hawking peak ($m<0$) the intensity goes with  $(|\widetilde{g}_-|^2+|\widetilde{g}_\mathrm{B}|^2/X^2)\,X^{-2m}$ while on the right ($m\ge0$, including the Hawking peak) with $(|\widetilde{g}_-|^2+|\widetilde{g}_\mathrm{B}|^2X^2)\,X^{2m}$. For zero modulation, the spectrum reduces to the Hawking peak at $\omega_-'=-\omega_2'$ and the backreaction at $\omega_-'-\delta'$. For non--zero modulation, the backreaction still appears in the asymmetry of the sidebands at $\omega_-+m\delta'$.  Given a measured spectrum, we may retrieve the Hawking--radiation and backreaction intensities from the data, as we show next. 

\setcounter{equation}{0}
\section{Data analysis}

First we discuss the dispersion data (Fig.~2). They are determined by measurements of the group velocity in the IR combined with modeling based on electron microscopy of the fibre cross section, as described previously$^9$. We noticed a systematic shift in the measured wavelength of Hawking radiation due to months of light exposure of the fibre and corrected for it by adding to $\beta=n_0\,\omega/c$ a small, constant $\delta\beta_0$ that does not change the group velocity. We determine $\delta\beta_0$ requiring that it reproduces both the measured $\omega_-$ and the measured horizon $\omega_h$ (with $\lambda_h=1551\,\mathrm{nm}$  for a $1450\,\mathrm{nm}$ probe$^9$). 

We take the UV data (Fig.~3) $I_{12}$ with both pump and probe present and the UV data $I_1$ without the probe, each $50\%$ of the time, in order to subtract the background of Negative--frequency Resonant Radiation (NRR)$^{30}$. Taking the difference between the two spectra should give us our signal. However, the probe perturbs the pump and hence the NRR, despite it being $10^{-2}$ less intense$^{35}$. The pump pulse fractures into solitons that, like black holes in General Relativity, are characterized by only a few parameters. These are the carrier frequency $\omega_0$ and the duration $\tau_0$. The intensity is locked to the duration as $\tau_0^{-2}$ and the group velocity is set by $\omega_0$. Both $\omega_0$ and $u$ define the NRR frequency in the dispersion diagram (Fig.~2).  So if the carrier frequency changes $\omega_\mathrm{NRR}$ changes (in the order of $10^{-3}\omega_\mathrm{NRR}$). The spectral shape of the NRR is less related to the pump spectrum and remains  intact. If we do not correct for the spectral shift of the NRR we may get negative values of the signal intensity$^9$. For performing the correction, we select in each data set a relevant range around the NRR peak, minimize the root mean square of $I_\mathrm{signal}=I_{12}-I_1-\delta\omega_0\,\mathrm{d}I_1/\mathrm{d}\omega$ for $\delta\omega_0$ and  take  $I_\mathrm{signal}$ for all wavelengths as our signal. 

The spectral shapes of  the Hawking and backreaction bands and their sidebands are identical, as the Fourier--transformed field $\widetilde{a} \propto \sum_m \mathrm{e}^{-\mathrm{i}\omega_m'\zeta} \widetilde{g}_m (-\mathrm{i}X)^m \widetilde{a}_\mathrm{H}(\omega-\omega_m)/\Delta]$ is given by the Fourier--transformed amplitude $\widetilde{a}_\mathrm{H}$ of the Hawking radiation (emitted with unity strength) with the coefficients $\widetilde{g}_m$ of Eq.~(\ref{eq:sum}). The peak frequencies $\omega_m$ are determined by $\omega_m'=\omega_-'+m\delta'$, which translates into the laboratory frequencies $\omega_m=\omega_--m\,(\omega_\mathrm{B}-\omega_-)$ as the dispersion curve (Fig.~2c) is nearly linear there. Due to variations in the pump, the propagation time $\zeta$ (in the order of $5\times 10^3 /\delta'$) varies more than $2\pi/\delta'$: we need to average. As a result, the intensity is an incoherent sum of the Hawking spectra, despite the bands being coherent:
%%%%%%
\begin{equation}
I_\mathrm{signal}\propto \sum_m |\widetilde{g}_m|^2 X^{2m} I_\mathrm{H}(\omega-\omega_m) \,.
\label{eq:fit}
\end{equation}
%%%%%%
The Hawking spectrum $I_\mathrm{H}(\omega)$ is emitted while the probe is being redshifted. We found empirically that it interpolates between the $\sech^2(\nu)$ of a soliton and a parabolic profile $1-\nu^2$ with $\nu=\omega/\Delta$ (and spectral widths $\Delta$) as $I_\mathrm{H} = 1-\theta^2$ with $\theta = \int_0^\nu \sech^2(\upsilon^\mu)\,\mathrm{d}\upsilon/\int_0^\infty \sech^2(\upsilon^\mu)\,\mathrm{d}\upsilon$. For a soliton we have $\mu=1$ and for a parabola $\mu\rightarrow\infty$. For our six data sets (Fig.~4) we found $\mu=(10.0, 1.2, 1.1, 10.0, 1.0, 1.05)$. We determine the parameters $\omega_-$, $\omega_\mathrm{B}$, $\Delta$, $X$ and the intensities $|\widetilde{g}_-|^2$ and $|\widetilde{g}_\mathrm{B}|^2$ by fitting. For each of the six data sets we determine $u/c$ in the Doppler formula (1) requiring that it reproduces the determined $\omega_-$. We need to do this due to the Raman shifting of the pump at the point of emission of Hawking radiation that is difficult to control. Then the $\omega_\mathrm{B}$ agree well with their predicted values from the dispersion data (Fig.~2).

The theoretical curve [Eq.~(\ref{eq:fit})] fits the UV data well (Fig.~4) except in the region of sideband $m=1$. There $\omega'$ coincides with the negative co--moving frequency of the pump, {\it i.e.} with the frequency of the NRR (Fig.~2). As the pump and the probe are coherent, the NRR and this sideband are also coherent and interfere. As they oscillate with the same frequency this interference is not averaged out, but remains visible in the data (Fig.~4). To avoid the interference, the  fitting is done only with data to the left and including the Hawking peak ($\lambda\lesssim\lambda_-$) but it still gives a remarkably good fit for sideband $m=2$ (Fig.~4a and 4d). Without our modified subtraction procedure for $I_\mathrm{signal}$ this sideband would be drowned in negative values of $I_{12}-I_1$. The fact that it appears and agrees with a fit done on the other end of the spectrum shows the consistency of the data analysis.

The NRR intensity varies greatly from day to day (up to a factor of 3) due to imperceptible variations in the incoupling efficiency and so does the intensity of the Hawking radiation. Yet they are related: NRR is Hawking radiation made by the pump pulse and stimulated by the pump itself. This implies that its intensity falls exponentially in frequency with the same exponent as Hawking radiation:
%%%%%%
\begin{equation}
I_\mathrm{NRR}=I_1\exp\left(-\frac{\omega_1}{\omega_\mathrm{H}}\right) \quad\mbox{and}\quad I_\mathrm{H}=I_+\exp\left(-\frac{\omega_+}{\omega_\mathrm{H}}\right)
\label{eq:NRR}
\end{equation}
%%%%%%
with constants $I_1$ and $I_+$. Here $\omega_1$ denotes the frequency of the pump, $\omega_+$ the redshifted frequency of the probe and $\omega_\mathrm{H}$ the effective frequency characterizing the Boltzmannian tail of the Planck spectrum of Hawking radiation in the limit of large frequencies. For each data set we apply our fitting procedure to extract the photon flux of the stimulated Hawking radiation $|\widetilde{g}_-|^2$ and divide it by the stimulant, the measured average coupled probe power $P$  ($0.66\mathrm{mW}$, $0.99\mathrm{mW}$, $1.35\mathrm{mW}$, $1.2 \mathrm{mW}$, $1.2 \mathrm{mW}$,  $0.81  \mathrm{mW}$) divided by the original $\omega_2$ (for converting energies $\hbar\omega N$ to photon numbers $N$) such that $I_+=\mathrm{const}$. The backreaction $I_\mathrm{B}$ should also go as $I_0\exp(-\omega_+/\omega_\mathrm{H})$ with another constant $I_0$ and be quadratic in the probe intensity. We obtain
%%%%%%
\begin{equation}
\frac{I_\mathrm{H}}{I_+} = \frac{I_\mathrm{B}}{I_0} = \left(\frac{I_\mathrm{NRR}}{I_1}\right)^{\omega_+/\omega_1} \quad \mbox{with}\quad I_\mathrm{H} = \frac{\omega_2}{P}\,|\widetilde{g}_-|^2 \quad\mbox{and}\quad I_\mathrm{B} = \left(\frac{\omega_2}{P}\right)^2\,|\widetilde{g}_\mathrm{B}|^2
\label{eq:power}
\end{equation}
%%%%%%
with the unknown constants $I_1$, $I_+$, $I_0$ and the known ratios $\omega_+/\omega_1=\lambda_1/\lambda_+$ for the six data sets (Fig.~4). The logarithms of $I_\mathrm{H}/I_\mathrm{NRR}^{\omega_+/\omega_1}$ and $I_\mathrm{B}/I_\mathrm{NRR}^{\omega_+/\omega_1}$ should lie on straight lines in $\omega_+/\omega_1$ with the same slope. In the linear fits we exclude the $1100\,\mathrm{nm}$ data. There the probe lies at the edge of the capture range where the red--shifted efficiency drops (we see this by comparing the signal with the NRR in Fig.~3). As we normalize with respect to the incident probe intensity (not the red--shifted one) we would otherwise make a systematic error. The other data points lie on straight lines (Fig.~5) with relative root--mean squares of the deviations (RMS divided by the signal) of $0.015$ for the Hawking radiation and $0.010$  for the backreaction. The slopes agree within $2\%$, {\it i.e.} within the combined error of the slopes. This shows the thermality of the Hawking--radiation  and backreaction spectra and also the fact that the backreaction goes quadratically with probe intensity. Figure 3c also shows that the experiment is run in a regime where both effects are of roughly equal magnitude ($I_+\sim I_0$).

\setcounter{equation}{0}
\section{Bose Einstein condensates and tetrads}

Here we show how the concept of a Hawking Hamiltonian applies to other analogues of gravity and possibly to gravity itself. The best--understood analogue of gravity is based on Bose--Einstein condensates. There the theory$^{38}$ starts from the grand--canonical Hamitonian for a Bose gas $\widehat{\psi}$ of atoms with mass $m$ confined by the trapping potential $U$ and interacting by point collisions with strength $g$: 
%%%%%%
\begin{equation}
\widehat{H}-\mu\widehat{N} = \int \widehat{\psi}^\dagger \left(-\frac{\hbar^2\nabla^2}{2m}+U-\mu+\frac{g}{2}\widehat{\psi}^\dagger\widehat{\psi}\right) \widehat{\psi}\,dV
\label{eq:grand}
\end{equation}
%%%%%%
where $\mu$ is the chemical potential (the energy per atom). The atoms shall be Bose--condensed with macroscopic wave function $\psi_0=\sqrt{\rho}\,\mathrm{e}^{\mathrm{i}\varphi}$. We split the quantum field $\widehat{\psi}$ into the condensate and the field $\widehat{\phi}$ of elementary excitations, $\widehat{\psi}=\psi_0+\mathrm{e}^{\mathrm{i}\varphi}\widehat{\phi}$, and expand the Hamiltonian (\ref{eq:grand}) up to quadratic order in $\widehat{\phi}$. We obtain three Hamiltonians,$\widehat{H}-\mu\widehat{N} \sim H_0 + \widehat{H}_1 + \widehat{H}_2$, the zeroth--order classical Hamiltonian $H_0$ that generates the Gross--Pitaevskii equation$^{38}$ of the condensate, the first--order Hamiltonian $\widehat{H}_1$ that vanishes under the Gross--Pitaevskii equation$^{38}$ and the quadratic Hamiltonian $\widehat{H}_2$ that generates the Bogoliubov--de Gennes equation $^{38}$ of the elementary excitations:
%%%%%%
\begin{equation}
\mathrm{i}\hbar\partial_t\,\widehat{\phi} = (T + U - \mu +2 mc^2)\widehat{\phi} + mc^2 \widehat{\phi}^\dagger \quad \mbox{where}\quad mc^2 = g\varrho
\label{eq:BdG}
\end{equation}
%%%%%%
with $c$ being the speed of sound (not light) and $T$ the kinetic energy $T=\frac{m}{2}(\frac{-\mathrm{i}\hbar \nabla}{m} +\bm{u})^2$ with $\bm{u}=\frac{\hbar\nabla\varphi}{m}$ as the flow of the condensate. Both $c$ and $\bm{u}$ may vary. The simplest way of seeing how the Bogoliubov--de Gennes dynamics represents an analogue of gravity is considering an excitation that is a modulated plane wave with wavevector $\bm{k}$ and frequency $\omega$. We put $U=0$ and $\mu=mc^2+\frac{m}{2}u^2$ and obtain from Eq.~(\ref{eq:BdG}) the dispersion relation $(\omega-\bm{u}\cdot\bm{k})^2=c^2k^2[1+k^2/(2k_0)^2]$ with $k_0=mc/\hbar$. Here $\omega-\bm{u}\cdot\bm{k}$ represents the Doppler shift due to the motion of the condensate --- the dispersion relation is simply Bogoliubov's dispersion relation$^{38}$ in a locally co--moving frame.  In the regime of $k\ll k_0$ we can give it a completely relativistic form:
%%%%%%
\begin{equation}
g^{\alpha\beta}k_\alpha k_\beta = 0 \quad\mbox{with}\quad g^{\alpha\beta} = c^{-2}\begin{pmatrix} 1 & \bm{u} \\ \bm{u} & -c^2 + \bm{u}\otimes\bm{u} \end{pmatrix} \quad\mbox{and}\quad k_\alpha = \left(-{\omega},\bm{k}\right)
\label{eq:reldispersion}
\end{equation}
%%%%%%
adopting Einstein's summation convention. To the inverse metric tensor  $g^{\alpha\beta}$ corresponds Unruh's famous acoustic metric$^{23}$:
%%%%%%
\begin{equation}
\mathrm{d}s^2=g_{\alpha\beta}\,\mathrm{d}x^\alpha \mathrm{d}x^\beta = (c^2-u^2)\,\mathrm{d}t^2+2\bm{u}\cdot\mathrm{d}t\,\mathrm{d}\bm{r}-\mathrm{d}\bm{r}^2 \,.
\label{eq:unruh}
\end{equation}
%%%%%%
The metric shows that, when the condensate reaches the speed of sound, time appears to stand still for the elementary excitations and so a horizon may be established. The Hamiltonian for the resulting Hawking radiation we read off from the grand--canonical Hamiltonian (\ref{eq:grand}) by splitting the atomic field $\widehat{\psi}$ into the condensate and the excitation as $\widehat{\psi}=\widehat{\psi}_0+\widehat{\psi}_1$ assuming now that the condensate is as quantum as the excitations (since it needs to provide the energy for Hawking radiation). We express the excitations in terms of the Bogoliubov-de Gennes modes $u_\nu$ and $v_\nu$ as $\widehat{\psi}_1=\sum_\nu\, (u_\nu\hat{a}_\nu+ v_\nu^*\,\widehat{a}_\nu^\dagger)$. We see that the grand--canonical Hamiltonian (\ref{eq:grand}) contains processes of the type $\widehat{H}_\mathrm{R}=g\widehat{\psi}_0^\dagger\widehat{\psi}_0(u_{\nu_1}^*v_{\nu_2}^*\widehat{a}_{\nu_1}^{\dagger}\widehat{a}_{\nu_2}^{\dagger} +u_{\nu_1} v_{\nu_2} \widehat{a}_{\nu_1} \widehat{a}_{\nu_2})$ completely analogous to our optical case. Energy conservation requires that one of the two modes must have the exact negative frequency of the other, which is the hallmark of Hawking radiation. The backaction of this Hamiltonian should be noticeable in Bose--Einstein condensates as well, but has not been observed yet. 

Other analogues of gravity share similar bi--quadratic interactions, and possibly gravity itself. This is because the central quantity of gravity is the metric tensor $g_{\alpha\beta}$ of the space--time geometry and one can always express $g_{\alpha\beta}$ in a quadratic form. This is done$^{45}$ in terms of tetrads $\bm{e}_\alpha$ as $g_{\alpha\beta}=\bm{e}_\alpha\cdot \bm{e}_\beta$ where the dot abbreviates the scalar product with respect to the Minkowski metric $\eta_{\mu\nu}=\mathrm{diag}(1,-1,-1,-1)$. The inverse metric tensor is expressed in terms of the inverse tetrads as $g^{\alpha\beta}=\bm{e}^\alpha\cdot \bm{e}^\beta$ where $\bm{e}^\alpha$ is the matrix inverse of $\bm{e}_\alpha$. To give a simple example of tetrads, for the Unruh metric (\ref{eq:unruh}) we have $\mathrm{d}s^2=c^2\mathrm{d}t^2-\mathrm{d}\bm{r}'^2$ with $\mathrm{d}\bm{r}'=\mathrm{d}\bm{r}-\bm{u}\,\mathrm{d}t$ and hence $\bm{e}_\alpha\,\mathrm{d}x^\alpha = (\mathrm{d}t,\mathrm{d}\bm{r}')$. We thus obtain
%%%%%%
\begin{equation}
\bm{e}_\alpha = \begin{pmatrix}c & \bm{0}\\ -\bm{u} & \mathbb{1} \end{pmatrix}\quad\mbox{and so}\quad \bm{e}^\alpha = \begin{pmatrix}1/c & \bm{u}/c\\ \bm{0} & \mathbb{1} \end{pmatrix} .
\label{eq:utetrad}
\end{equation}
%%%%%%
In the tetrad formalism$^{45}$, the interaction of gravity with other fields is typically bi--quadratic. To see this, consider a scalar field $\varphi$. The interaction Lagrangian with gravity is proportional to $g^{\alpha\beta}(\partial_\alpha\varphi)(\partial_\beta\varphi)=\bm{e}^\alpha\cdot \bm{e}^\beta(\partial_\alpha\varphi)(\partial_\beta\varphi)$ which is bi--quadratic.  Einsteinian gravity is not a renormalizable quantum field theory (as the Einstein--Hilbert action$^{45}$ depends on second derivatives of the metric tensor and hence goes with the inverse second power of a length, which blows up in a perturbative expansion). Yet it could establish a low--energy effective field theory. There are several options for such a theory. Our analysis of Hawking radiation suggests that it might be wise to take the tetrad formalism as the starting point of an effective theory of quantum gravity.

\section*{Data availability}

Data is available upon request. 

\section*{Additional references}

\begin{enumerate}
\setcounter{enumi}{50}
		
\item
Couairon, A. et. al. Practitioner's guide to laser pulse propagation models and simulation. Europhys. J. Special Topics, \textbf{199}, 5--76 (2011).

\item
Akhmediev, N. \& Karlsson, M.
Cherenkov radiation emitted by solitons in optical fibers.
Phys. Rev. A, \textbf{51}, 2602--2607, (1995).

\item
Dudley, J. M. \& Taylor, J. R. \textit{Supercontinuum generation in optical fibers}. (Cambridge University Press, Cambridge, 2010).
	
\end{enumerate}

\section*{Acknowledgements}

We are grateful to 
S. Amiranashvili,
U. Bandelow,
H. Cohen,
J. Drori,
M. Gelvan,
F. K\"{o}nig,
Y. Rosenberg,
S. Rotter,
and most of all to the late Y. Silberberg
for help, advice and scientific discussions.

\section*{Funding}
L.M.P. and U.L. were supported by the Israel Science Foundation, the Murray B. Koffler Professorial Chair and (U.L) by a Global Fellowship of the Vienna Institute of Technology. R.A-S. and D.B. acknowledge the support of Conahcyt (Mexico) Ciencia de Frontera 51458-2019 and the Marcos Moshinsky Chair (2023).

\section*{Author contributions}
L.M.P. was conducting, analysing and leading the experiment, R.A.-S. and D.B. were performing numerical simulations, U.L. developed the analytical theory, D.B. and U.L. did the data analysis, supervised the project, and conceived the idea of this paper.

\section*{Competing interests}

There are no competing interests.

\section*{Additional information}

Correspondence and requests for materials should be addressed to Ulf Leonhardt by email\\
ulf.leonhardt@weizmann.ac.il.

\end{document}